\documentstyle[12pt]{article}
\begin{document}
\begin{flushright}
IP/BBSR/95-25 \\
nucl-th/9503029 \\
\end{flushright}
\begin{center}
{\bf IMPLICATIONS OF THEORETICAL IDEAS REGARDING COLD FUSION}
\vspace{1in}
{\newline\bf AFSAR ABBAS}
{\newline Institute of Physics}
{\newline Bhubaneswar - 751005, India}
\vspace{1.5in}
{\newline\bf Abstract}
\end{center}

A lot of theoretical ideas have been floated to explain the so called
cold fusion phenomenon. I look at a large subset of these and study
further physical implications of the concepts involved.
I suggest that these can be
tested by other independent physical means. Because of the
significance of these the experimentalists are urged to look for
these signatures. The results in turn will be important for a better
understanding and hence control of the cold fusion phenomenon.

\pagebreak

Since the initial claims, counterclaims and confusion of 1989 the
field of " cold fusion " has settled down as a reasonably well
pursued field all over the world as evidenced by several
recent  conferences
and publications [1-7]. Perhaps not surprisingly it has turned out to
be a tough field experimentally as much as the results viewed globally
are quite sporadic and the optimum conditions are still unknown.
However the bottomline is that whether conventional cold fusion or not
excess heat and/or neutron and/or $He^4$ etc are being produced
often in uncontrolled and unknown ways.

The theoretical situation regarding the cold fusion phenomenon
is no better than the experimental situation described above.
As many ideas as practically the number of people caring to
work in this field have sprung up ( see ref. [ 5, 8-10 ]  for
review of some of these). Several of these are in direct conflict
with each other. One would like to bring some order in these to
facilitate further understanding. A consistency study of a major
subset of these theoretical ideas is the purpose of this paper.

The end result or a major input or requirement of a large number
of theoretical papers to explain the cold fusion phenomenon
( as a sample and review see [5,8-10] ) is the following:

(a) The average separation of the deutrons in the $D_2$ molecule
under the conditions prevalent in the cold fusion experiments
decreases ( by say a factor of five ) from the equilibrium
distance of 0.74 A.

(b) The average distance between the electron and the D nuclei
decreases from the free space value of 0.5 A under the cold
fusion conditions.

(c) Both (a) and (b) are inseparably mixed up.

Given a model which falls in one of the above categories what more
can one say about it? Certainly such interesting situation as stated
in (a),(b) and (c)
besides leading to an enhanced rate of fusion at low temperatures
should have other physical implications. As we shall see it is indeed
so. Below I shall suggest further  experiments which should be able
to say whether one of the above is taking place or not and hence
should have implications for a better understanding  and control of
the cold fusion phenomenon.

Let us look at a linear molecule ( with $D_2$ molecule as a
prototype ). Let the nuclear spin be $I^\rightarrow$ and the
molecular rotational spin be $J^\rightarrow$. With
$F^\rightarrow$=$I^\rightarrow$+$J^\rightarrow$ one finds that [11]
the quadrupole interaction energy is given by
\begin{displaymath}
W_{E2}={{e^2 q_J Q}\over {2I(2I-1)(2J+3)(2J-1)}}{[C(C+1){3\over 8}-
      { I(I+1)J(J+1)\over 2}]}
\end{displaymath}
where
\begin{displaymath}
C= F(F+1) - I(I+1) - J(J+1)
\end{displaymath}

Here Q is the quadrupole moment of the nucleus and $q_J$ is the
molecular field gradient along the symmetry axis. In general it is
not easy to determine $q_J$ for a molecule but  it can be easily
done for $D_2$ [12,13]. For $D_2$ it is given as
\begin{displaymath}
q= {-2J q^1 \over {(2J+3)}}
\end{displaymath}

where
\begin{displaymath}
q^1= {{1\over {R^3}}-{1\over 2} \int d^3r \rho (R,r,\theta)
    {{3 \cos ^2 {\theta}    -1}\over r^3}}
\end{displaymath}

Here R is the internuclear distance, $\rho$ is the elctron charge
density, r the radius vector from a nucleus to the charge and
$\theta$ is the angle that r
makes with the internuclear line.
The average has to be taken over the lowest molecular vibrational
states.
$q_J$ can be obtained for $D_2$ in free  space [12,13].
The quantity $e^2 q_J Q$ is called the quadrupole interaction
constant and can be determined experimentally very precisely [11].
Historically by putting in the value of $q_J$ the value of the
quadrupole moment Q of deuteron was discovered the first time this way
[14].

Now let us reverse the argument. Let the quadrupole interaction
energy for $D_2$ molecule be determined experimentally under the
conditions prevalent in the cold fusion [ 1-7,10]. As stated
earlier we expect to detect major differences from the free $D_2$
value. Let us make a reasonable assumption that the value of Q is
not affected. So any change in the quadrupole interaction constant
will be attributed to change in $q_J$.
One may then view this as a  manifestation of categories (a), (b)
and (c) as  discussed earlier. After the experimental information
is available then it may turn out that it is reasonable and
fruitful to view it as entirely due to only one of the categories
of either (a) or (b). This will quite clearly make the analysis much
easier. However it is more likely that it is category (c) which has
to be invoked to understand the data. This means that the aspects
(a) and (b) may not be isolated by a study of the quadrupole
interaction energy.

To isolate the effects given in (a) and (b) individually in addition
to the determination of the quadrupole interaction energy one has to
suggest other experiments to do the job. In principle there may be
several possibilities.
The magnetic hyperfine interaction may be used to do this [11].
However the effect visualised in (a) may be most easily  isolated
by studying the change in rotational spectra of $D_2$ under the
cold fusion conditions [11].

In summary, above I have isolated interesting theoretical aspects
of the cold fusion phenomenon. They are
simple and have implications which can be experimentally tested
through the study of hyperfine spectroscopy. Large effects are
expected under the conditions prevalent during the
cold fusion . I believe a proper experimental study
as suggested here will help in clearing the cold fusion situation,
both experimentally and theoretically.

\pagebreak

\begin{noindent}
{\bf References}
\vspace{0.5cm}
\newline [1] M. Fleishman , S. Pons, Phys. Lett {\bf A176} (1993) 118
\newline [2] S. Pons , M. Fleishman, Nuovo Cim. {\bf 105A} (1992) 763
\newline [3] A. B. Karabut, Ya. R. Kucherov , I. B. Savvatunova,
             Phys. Lett {\bf A170} (1992) 265
\newline [4] E. Botta, T. Bressani, D. Calvo, A. Feliciello,
           P. Gianotti, C. Lamberti, M. Angelo, F. Iazzi,
	  B. Minetti, A. Zecchina, Nouvo Cim. {\bf 105A} (1992) 1663
\newline [5] M. Srinivasan, Currrent Sc. {\bf 60} (1991) 417
\newline [6] P. K. Iyengar, M. Srinivasan et al,
           Fusion Tech. {\bf 19} (1990) 32
\newline [7] A. Takahashi, T. Iida, F. Maekawa, H. Sugimoto ,
           S. Yoshida, Fusion Tech. {\bf 19} (1991) 380
\newline [8] T. Bressani, E. Del Giudice , G. Preparata,
           Nouvo Cim. {\bf 101A} (1989) 845
\newline [9] G. Preparata, University of Milano preprint: MITH 93/7
\newline [10] V. A. Tsarev, Sov. Phys. Usp. {\bf 10} (1992) 842
\newline [11] C. H. Townes , A. L. Schawlow, " Microwave
           Spectroscopy" Dover Pub. , New York, 1975
\newline [12] A. Nordsieck, Phys. Rev. {\bf 58} (1940) 310
\newline [13] G. F. Newell, Phys. Rev. {\bf 78} (1950) 711
\newline [14] J. M. B. Kellog, I. I. Rabi, N. F. Ramsey ,
           J. R. Zacharias, Phys. Rev. {\bf 55} (1939) 318
\end{noindent}
\end{document}